\documentclass[11pt,a4paper]{article}

\usepackage[utf8]{inputenc}
\usepackage[T1]{fontenc}
\usepackage{lmodern}
\usepackage[a4paper,margin=2.5cm]{geometry}
\usepackage{microtype}
\usepackage[english]{babel}
\usepackage{csquotes}
\usepackage[table]{xcolor}
\usepackage{amsmath}
\usepackage{amssymb}
\usepackage{graphicx}
\usepackage{booktabs}
\usepackage{array}
\usepackage{tabularx}
\usepackage{ragged2e}
\usepackage{enumitem}
\usepackage{tikz}
\usepackage{xurl}
\usepackage[hidelinks]{hyperref}
\usepackage[font=small,labelfont=bf]{caption}
\usepackage{stfloats}
\usepackage{placeins}
\usepackage{titlesec}
\usepackage{longtable}
\usepackage{xltabular}
\usepackage{float}
\usepackage{colortbl}
\usepackage[hidelinks]{hyperref}

\definecolor{aiBlueDark}{HTML}{183A66}
\definecolor{aiBlueFill}{HTML}{EEF5FC}
\definecolor{aiGreenDark}{HTML}{2F6B3C}
\definecolor{aiGreenFill}{HTML}{EEF7EE}
\definecolor{aiPurpleDark}{HTML}{5D3F8C}
\definecolor{aiPurpleFill}{HTML}{F3EFFA}
\definecolor{aiOrangeDark}{HTML}{B45A1C}
\definecolor{aiOrangeFill}{HTML}{FFF4EA}
\definecolor{aiGrayDark}{HTML}{4A5568}
\definecolor{aiGrayFill}{HTML}{F3F5F7}
\definecolor{linkGray}{HTML}{C7CCD3}

\titlespacing*{\section}{0pt}{16pt plus 2pt minus 1pt}{12pt plus 2pt minus 1pt}
\titlespacing*{\subsection}{0pt}{12pt plus 2pt minus 1pt}{8pt plus 1pt minus 1pt}
\titlespacing*{\subsubsection}{0pt}{10pt plus 1pt minus 1pt}{6pt plus 1pt minus 1pt}


\renewcommand{\arraystretch}{1.18}

\newcolumntype{L}[1]{>{\RaggedRight\arraybackslash}p{#1}}
\newcolumntype{Y}{>{\RaggedRight\arraybackslash}X}
\newcolumntype{Z}{>{\justifying\arraybackslash}X}
\newcolumntype{J}[1]{>{\justifying\arraybackslash}p{#1}}
\newcolumntype{C}[1]{>{\Centering\arraybackslash}p{#1}}

\usetikzlibrary{positioning, arrows.meta, shapes.geometric, calc, fit, backgrounds, decorations.pathmorphing, decorations.pathreplacing}

\tikzset{
  boxBlue/.style={rectangle, rounded corners=3pt, draw=aiBlueDark, fill=aiBlueFill, align=center, minimum height=9mm, text width=27mm, font=\scriptsize},
  boxGreen/.style={rectangle, rounded corners=3pt, draw=aiGreenDark, fill=aiGreenFill, align=center, minimum height=9mm, text width=27mm, font=\scriptsize},
  boxPurple/.style={rectangle, rounded corners=3pt, draw=aiPurpleDark, fill=aiPurpleFill, align=center, minimum height=9mm, text width=27mm, font=\scriptsize},
  boxOrange/.style={rectangle, rounded corners=3pt, draw=aiOrangeDark, fill=aiOrangeFill, align=center, minimum height=9mm, text width=27mm, font=\scriptsize},
  boxGray/.style={rectangle, rounded corners=3pt, draw=aiGrayDark, fill=aiGrayFill, align=center, minimum height=9mm, text width=29mm, font=\scriptsize},
  db/.style={cylinder, shape border rotate=90, aspect=0.25, draw=aiPurpleDark, fill=aiPurpleFill, align=center, minimum height=11mm, text width=27mm, font=\scriptsize},
  decision/.style={diamond, draw=aiOrangeDark, fill=aiOrangeFill, aspect=2.2, align=center, text width=25mm, font=\scriptsize, inner sep=1pt},
  arrow/.style={-{Latex[length=2.2mm]}, line width=0.5pt, draw=aiGrayDark},
  dashedarrow/.style={-{Latex[length=2.2mm]}, dashed, line width=0.45pt, draw=aiGrayDark},
  group/.style={draw=aiGrayDark, rounded corners=2pt, inner sep=5pt, dashed},
  timeline/.style={line width=0.6pt, draw=aiGrayDark, -{Latex[length=2.2mm]}},
  event/.style={circle, draw=aiBlueDark, fill=aiBlueFill, minimum size=3mm, inner sep=0pt}
}

\title{The Distributed Open-Source Vulnerability Ecosystem}

\author{
Peter Mandl\\
Munich University of Applied Sciences\\
Munich, Germany\\
\texttt{peter.mandl@hm.edu}
\and
Paul Mandl\\
Findustrial GmbH\\
Schörfling am Attersee, Austria\\
\texttt{paul.mandl@findustrial.io}
}

\date{}

\begin{document}

\maketitle

% -----------------------------------------------------------------------
% ABSTRACT
% -----------------------------------------------------------------------

\begin{abstract}

Identifying known software vulnerabilities is a central task in software
supply chain security management. Although publicly available vulnerability
information is based on shared standards, different vulnerability scanners
often report divergent results for identical software inventories. These
differences do not arise solely from individual data sources or scanner
implementations. They can emerge at several stages of the open-source
vulnerability ecosystem.

This paper presents a conceptual framework that describes vulnerability
management as a distributed process of information exchange and
transformation. It traces vulnerability information from its creation and
standardization through enrichment to context-dependent interpretation. The
analysis identifies heterogeneous information sources, divergent identity
and version models, temporal change, and context-dependent assessment as
major causes of inconsistent scanner findings. It then discusses the
implications for interpreting analysis results, designing reproducible
evaluation methods, and handling dynamic vulnerability knowledge in
practice.

\end{abstract}


%%%%%%%%%%%%%%%%%%%%%%%%%%%%%%%%%%%%%%%%%%%%%%%%%%%%%%%%%%%%%%%%%%%%%%%%%%%%%%%%
\section{Introduction}
\label{sec:introduction}
%%%%%%%%%%%%%%%%%%%%%%%%%%%%%%%%%%%%%%%%%%%%%%%%%%%%%%%%%%%%%%%%%%%%%%%%%%%%%%%%

Free and open-source software underpins much of modern software. Applications
often consist of numerous external libraries and frameworks distributed
through independent package ecosystems and maintained by separate projects.
This reuse shortens development cycles but also creates complex direct and
transitive dependencies. A vulnerability in a single component can therefore
propagate to many dependent applications
\cite{pashchenko2018vulnerable,ponta2020detection}. Studies of the npm, RubyGems, and Maven ecosystems have shown that transitive dependencies can substantially increase the exposure of software systems to known vulnerabilities \cite{zerouali2022impact,mir2023transitivity}.

Identifying such vulnerabilities is a central task in vulnerability management
for modern software supply chains. \emph{Software Composition Analysis} (SCA)
tools identify the components contained in a software artifact and compare
them with publicly available vulnerability information. Comparative studies
show, however, that different tools can produce substantially different
results even for comparable analysis targets
\cite{imtiaz2021comparative,zhao2023sca}. The differences concern not only the
number of reported vulnerabilities, but also component detection and mapping.
Assessments of relevant version ranges and assignments to specific packages
may likewise differ \cite{mandl2026groundtruth,wu2024affected}. Such
discrepancies are often attributed to individual databases or tool
implementations. In many cases, however, they arise earlier because the
underlying information is created and repeatedly transformed within a
distributed ecosystem.

From its initial discovery to its application in vulnerability management, vulnerability information passes through a series of organizational and technical processing stages. Throughout this process, it is coordinated, published, validated, and enriched by multiple information sources. It
is also transferred between different identity models and interpreted within
a specific application context. Available vulnerability knowledge therefore
does not originate from a single source, but results from the interaction of
independent actors and information systems. The literature examines many
parts of this process. Existing work addresses inconsistencies in public
vulnerability reports \cite{dong2019viem}, the linking of heterogeneous
information \cite{qin2023vulnerability}, and the construction of broad
vulnerability datasets \cite{ruan2024vulzoo}. Further studies investigate the
mapping of software components to vulnerability records
\cite{wu2024affected} and the limitations of current SCA methods
\cite{dietrich2024blindspots}. Less attention has been paid to how these
parts interact within a common information flow and how that interaction
produces divergent analysis results.

\begin{figure*}[t]
    \centering
    \includegraphics[width=\textwidth]{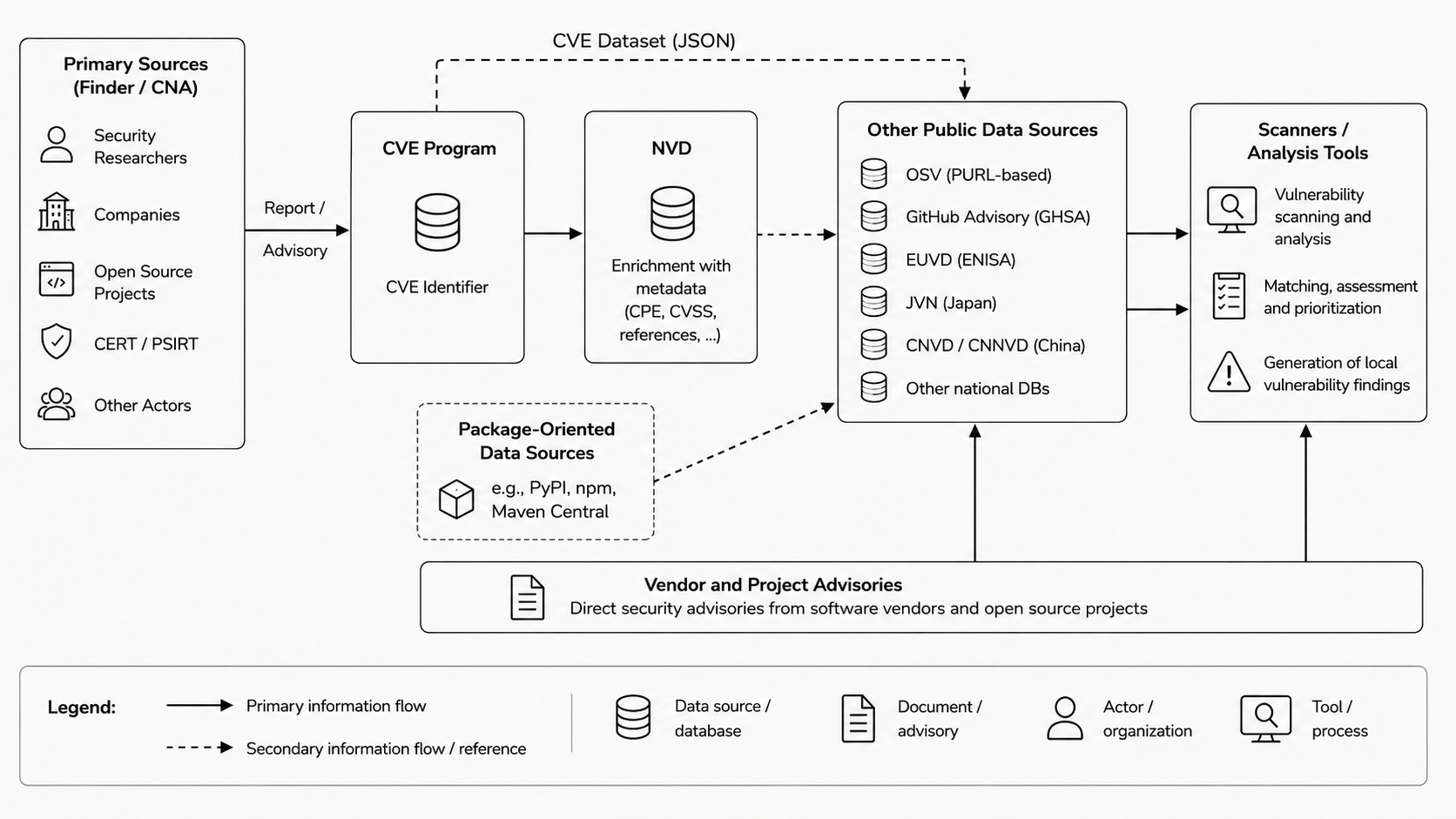}
    \caption{Information flow and key actors in the open-source vulnerability
    ecosystem. The flow extends from discovery and coordinated disclosure
    through enrichment in public information systems to the mapping of
    vulnerability information to software components. The transitions mark
    points at which separate technical decisions are made.}
    \label{fig:figure_1}
\end{figure*}

This paper examines the open-source vulnerability ecosystem as a connected
information process. Its focus is not on individual databases or analysis
tools, but on the creation, transformation, and interpretation of
vulnerability information. Figure~\ref{fig:figure_1} outlines the processing
stages discussed below. The flow begins with the discovery and reporting of a
vulnerability. Once a \emph{Common Vulnerabilities and Exposures} (CVE) identifier has been assigned, the published record
is taken up by different information systems, enriched, and linked to further
sources. Alongside this primary flow, secondary relationships connect package
data, vendor and project advisories, and other public databases. Scanners and
analysis tools combine these inputs in the local application context. The
figure also shows that separate technical decisions are made at several
transitions. These decisions can later produce divergent analysis results.
The figure therefore serves both as an overview and as a reference point for
the argument developed throughout the paper.

The remainder of the paper is organized as follows. Section~\ref{sec:ecosystem}
explains how vulnerability information is created, integrated, and
interpreted. Section~\ref{sec:discrepancies} examines the main causes of
divergent analysis results, including heterogeneous information sources,
different identity and version models, temporal change, and context-dependent
assessment. Section~\ref{sec:implications} discusses the implications for
research and practice. The final section summarizes the main findings.

%%%%%%%%%%%%%%%%%%%%%%%%%%%%%%%%%%%%%%%%%%%%%%%%%%%%%%%%%%%%%%%%%%%%%%%%%%%%%%%%
\section{The Open-Source Vulnerability Ecosystem}
\label{sec:ecosystem}
%%%%%%%%%%%%%%%%%%%%%%%%%%%%%%%%%%%%%%%%%%%%%%%%%%%%%%%%%%%%%%%%%%%%%%%%%%%%%%%%

The processing steps shown in Figure~\ref{fig:figure_1} are discussed below
along the flow of information. The process begins with the emergence of
information about a vulnerability. This information is subsequently
integrated into other information systems and interpreted in relation to a
specific application context.

%%%%%%%%%%%%%%%%%%%%%%%%%%%%%%%%%%%%%%%%%%%%%%%%%%%%%%%%%%%%%%%%%%%%%%%%%%%%%%%%
\subsection{Origin of Vulnerability Information}
\label{subsec:origination}
%%%%%%%%%%%%%%%%%%%%%%%%%%%%%%%%%%%%%%%%%%%%%%%%%%%%%%%%%%%%%%%%%%%%%%%%%%%%%%%%

The information flow begins with the discovery of a previously unknown
vulnerability. Relevant reports may originate from security research,
open-source projects, or operational practice. Following an initial report,
the technical cause and possible consequences are examined. It must also be
determined which products or versions are involved. Disclosure is often
coordinated so that the responsible projects or vendors can develop
appropriate countermeasures before publication
\cite{certcc2024disclosure}. Such processes are frequently supported by
Product Security Incident Response Teams (PSIRTs). Their tasks range from
receiving and assessing a report to coordinating remediation and publishing
security advisories \cite{first2024psirt}. Examples include Siemens
ProductCERT, which investigates vulnerability reports concerning Siemens
products and publishes corresponding security advisories; the Microsoft
Security Response Center (MSRC), which handles reports concerning Microsoft
products and services and coordinates their disclosure; and the Red Hat
Product Security Incident Response Team, which coordinates vulnerability
and incident response management for the Red Hat portfolio
\cite{siemens2026productcert,microsoft2026msrc,redhat2026psirt}. Different
assessments may already arise at this stage because technical information
can initially be incomplete and several parties may be involved in its
validation.

A confirmed vulnerability must be assigned an unambiguous reference for
information to be exchanged across organizational boundaries. The
CVE system provides globally
unique identifiers for this purpose \cite{cve2026}. These identifiers are
assigned and published by authorized \emph{CVE Numbering Authorities}
(CNAs), each of which is responsible for defined products, projects, or
other areas of scope \cite{cveprogram2026cna}. The Apache Software
Foundation is one example. As a CNA, it records vulnerabilities in the
projects under its responsibility and can assign the corresponding CVE
identifiers.

\begin{figure*}[t]
    \centering
    \includegraphics[width=\textwidth]{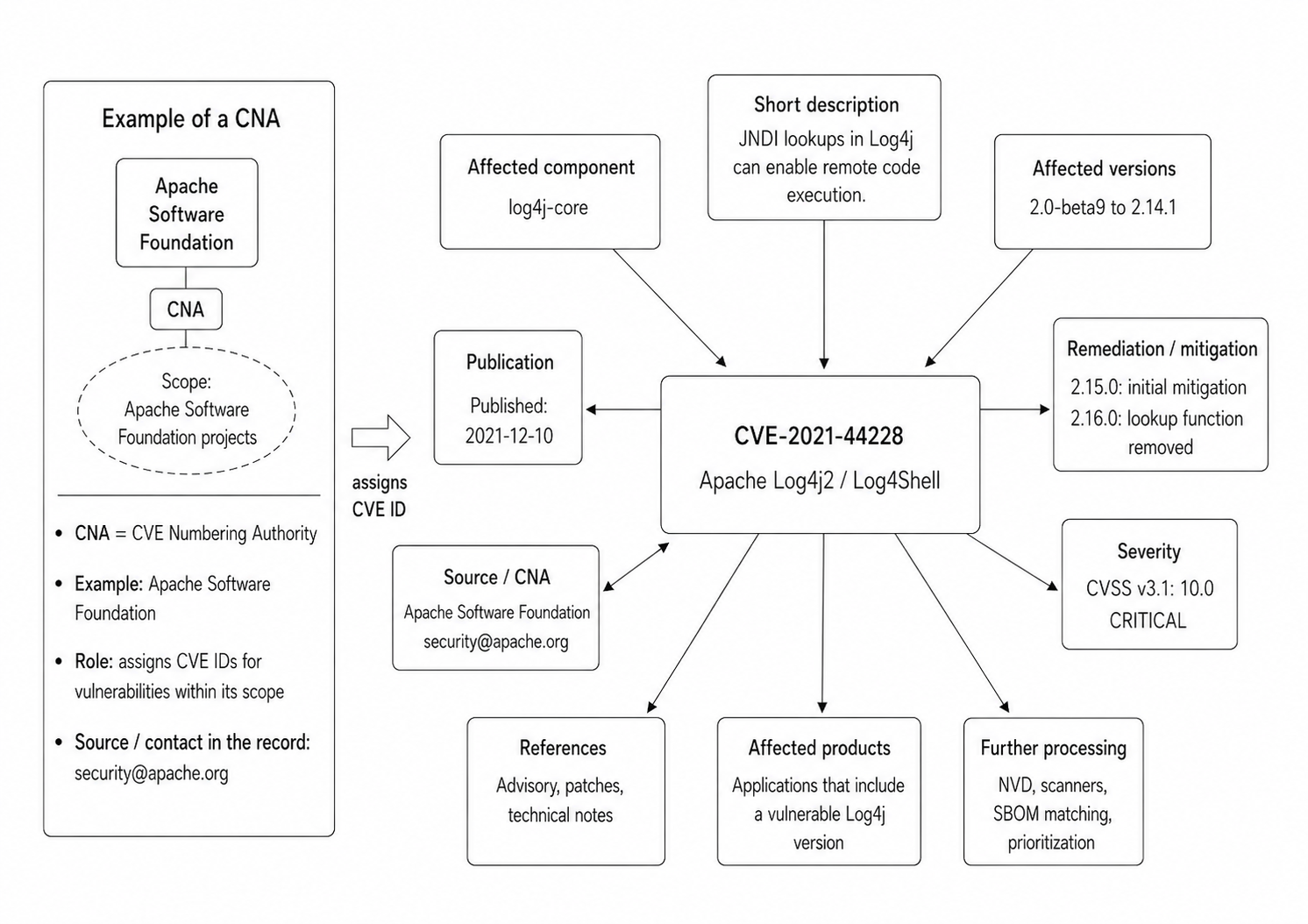}
    \caption{Schematic representation of a CVE Numbering Authority and the
    information associated with a CVE record, using CVE-2021-44228
    (\emph{Log4Shell}) as an example. The Apache Software Foundation acts
    as the responsible CNA, while downstream systems add further
    assessments and mappings to the published record.}
    \label{fig:cve-log4shell-informationen}
\end{figure*}

A CVE record provides the common reference point for subsequent processing.
It contains at least the information required for identification and
publication, although it may include additional structured information
depending on the responsible CNA. Typical elements include a description of
the vulnerability, information about the relevant product and versions, and
references to vendor advisories, patches, or further technical information
\cite{cve2026record}. Figure~\ref{fig:cve-log4shell-informationen}
illustrates these relationships using CVE-2021-44228
(\emph{Log4Shell}) as an example. It also shows that a CVE record cannot be
considered in isolation. Information supplied by the CNA is adopted by
downstream information systems, supplemented, and prepared for different
purposes.

The \emph{National Vulnerability Database} (NVD) plays a central role in
this subsequent enrichment by adding structured metadata to published CVE
records \cite{nvd2022process}. This metadata includes ratings based on the
\emph{Common Vulnerability Scoring System} (CVSS) \cite{cvss40},
mappings to the \emph{Common Weakness Enumeration} (CWE)
\cite{cwe2026}, and product and platform references based on the
\emph{Common Platform Enumeration} (CPE) \cite{cpe2026}. A common CVE
reference can therefore give rise to several interconnected and differently
enriched representations. These representations form the basis for further
processing of vulnerability information.

%%%%%%%%%%%%%%%%%%%%%%%%%%%%%%%%%%%%%%%%%%%%%%%%%%%%%%%%%%%%%%%%%%%%%%%%%%%%%%%%
\subsection{Integration and Transformation of Vulnerability Information}
\label{subsec:integration}
%%%%%%%%%%%%%%%%%%%%%%%%%%%%%%%%%%%%%%%%%%%%%%%%%%%%%%%%%%%%%%%%%%%%%%%%%%%%%%%%

The information flow does not end with the publication of a CVE record.
Other information systems adopt the available information, supplement it,
and adapt it to their respective purposes. A concise vulnerability
description is therefore transformed into several representations that
provide different technical perspectives on the same issue. While the
\emph{National Vulnerability Database} mainly enriches a CVE record with
standardized metadata and product references, other sources focus on
published software packages, particular package ecosystems, or national
coordination processes.

The \emph{Open Source Vulnerability Format} (OSV) uses a package-oriented
data model and describes affected versions within specific package
ecosystems \cite{osv2026schema}. The GitHub Advisory Database provides
comparable information for GitHub and supported package managers
\cite{ghsa2026}. Regional information systems supplement this global body
of knowledge through their own coordination, publication, and assessment
processes. These include the \emph{European Vulnerability Database} (EUVD)
\cite{enisa2025euvd}, \emph{Japan Vulnerability Notes} (JVN) and JVN
iPedia \cite{jvn2026}, as well as the Chinese databases
\emph{China National Vulnerability Database} (CNVD) \cite{cnvd2026} and
\emph{China National Information Security Vulnerability Database} (CNNVD)
\cite{cnnvd2026}. Vendor and project advisories remain important primary
sources because they often provide precise information about affected
versions, backports, workarounds, and available fixes
\cite{certcc2024disclosure,redhat2026backport}.

\begin{table*}[t]
    \centering
    \caption{Key information sources in the open-source vulnerability ecosystem.}
    \label{tab:vuln_sources}

    \begingroup
    \footnotesize
    \setlength{\tabcolsep}{4pt}
    \renewcommand{\arraystretch}{1.15}

    \rowcolors{2}{gray!6}{white}

    \begin{tabularx}{\textwidth}{L{30mm}L{48mm}Y}
        \toprule
        \rowcolor{gray!18}
        \textbf{Source} &
        \textbf{Identity} &
        \textbf{Contribution and focus} \\
        \midrule

        CVE &
        CVE identifier &
        Global identification system for publicly known vulnerabilities.
        CVE provides a common reference across sources but does not
        necessarily include a complete mapping to specific software
        packages and versions
        \cite{cve2026,cve2026record}. \\

        NVD &
        CVE identifier and CPE names &
        Enriches CVE records with structured metadata such as CVSS scores,
        CWE mappings, references, and product-oriented CPE configurations
        \cite{nvd2022process,cvss40,cwe2026,cpe2026}. \\

        OSV &
        OSV identifier, CVE and other aliases, package ecosystem,
        package name, and version ranges &
        Provides a package-oriented representation of vulnerabilities in
        open-source ecosystems. OSV describes affected and fixed versions
        according to ecosystem-specific version semantics
        \cite{osv2026schema}. \\

        GitHub Advisory Database &
        GHSA identifier, CVE aliases, package ecosystem, package name,
        and version ranges &
        Provides security advisories for GitHub projects and supported
        package ecosystems. The information is used by Dependabot and
        other GitHub security functions
        \cite{ghsa2026}. \\

        PSIRT, vendor, and project advisories &
        Vendor- or project-specific advisory identifiers, product and
        version names, and CVE aliases &
        Provides information close to the original source about
        vulnerabilities in the vendor's or project's own products.
        Such advisories often contain precise details about affected
        versions, backports, workarounds, and available fixes
        \cite{first2024psirt,certcc2024disclosure,redhat2026backport}. \\

        EUVD &
        EUVD identifier, CVE references, and product- or service-related
        information &
        European vulnerability database containing information about
        vulnerabilities, affected products, exploitation, and available
        countermeasures
        \cite{enisa2025euvd}. \\

        JVN and JVN iPedia &
        JVN or JVNDB identifiers, CVE references, and product information &
        Japanese information and coordination source for vulnerabilities.
        JVN publishes coordinated advisories, while JVN iPedia provides
        the corresponding information in database form
        \cite{jvn2026}. \\

        CNVD &
        CNVD identifier, CVE references, and product information &
        Chinese national vulnerability database with its own collection,
        assessment, and publication processes
        \cite{cnvd2026}. \\

        CNNVD &
        CNNVD identifier, CVE references, and product information &
        Another Chinese national vulnerability database with its own
        institutional setting and its own coordination and publication
        processes
        \cite{cnnvd2026}. \\

        \bottomrule
    \end{tabularx}

    \endgroup
\end{table*}

Table~\ref{tab:vuln_sources} summarizes the information sources considered, the identity models they use, and their respective contribution and technical focus. In the \emph{Identity} column, identity denotes the mechanisms used by an information system to uniquely identify software entities and relate them to vulnerability information. The sources differ not only in the metadata they provide but also in the underlying identity models on which they are based.
Further ambiguities arise from project renaming, forks, and
distribution-specific variants. Backports make version information
particularly difficult to interpret. Security fixes may be applied to older
versions without fundamentally changing their original version designation
\cite{redhat2026backport}. The integration of vulnerability information is
therefore not a simple copying process but a continuing transformation
between different information and identity models. The following subsection
discusses how this information is interpreted for a specific software
artifact.

%%%%%%%%%%%%%%%%%%%%%%%%%%%%%%%%%%%%%%%%%%%%%%%%%%%%%%%%%%%%%%%%%%%%%%%%%%%%%%%%
\subsection{Interpretation of Vulnerability Information}
\label{subsec:interpretation}
%%%%%%%%%%%%%%%%%%%%%%%%%%%%%%%%%%%%%%%%%%%%%%%%%%%%%%%%%%%%%%%%%%%%%%%%%%%%%%%%

Vulnerability information available in public information systems initially
has no direct relation to a specific software system. A local analysis is
required to connect it to the software components actually in use and place
it within an application-specific context. This step marks the transition
from generally available vulnerability knowledge to a concrete security
assessment.

\begin{figure*}[t]
    \centering
    \includegraphics[width=\textwidth]{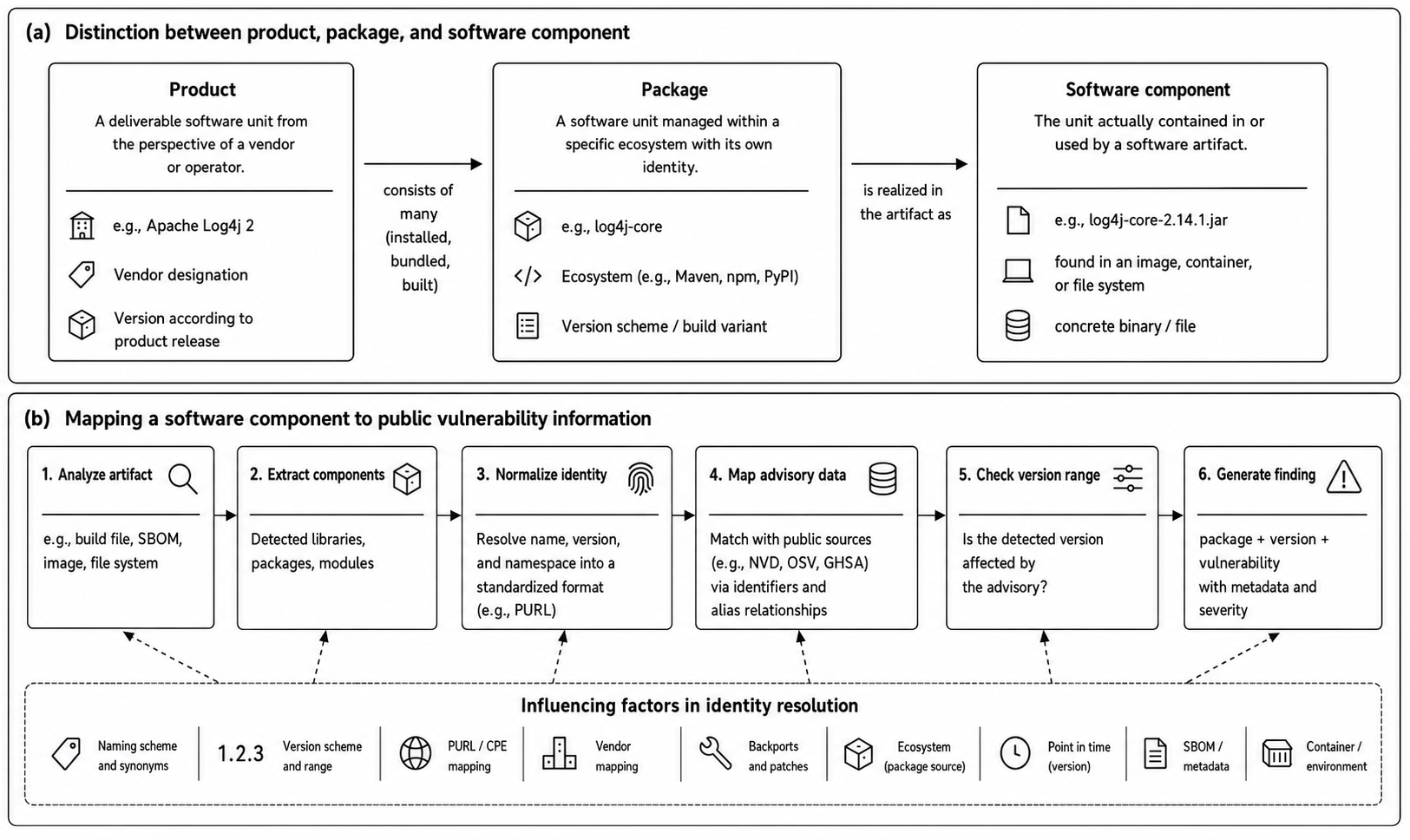}
    \caption{Semantic mapping between software components in use and public
    vulnerability information.}
    \label{fig:figure_3}
\end{figure*}

The software components contained in a software artifact must first be
identified. Their names and versions are then normalized, mapped to suitable
records in public information systems, and compared with the version ranges
specified in those records. Figure~\ref{fig:figure_3} illustrates this
process from the analysis of a software artifact to the generation of a
local vulnerability finding.

The mapping cannot rely solely on a name or version number. Public
information systems use different identity models to represent software and
vulnerabilities. Product-oriented sources commonly rely on product
identifiers, whereas package-oriented sources use package identities such
as the \emph{Package URL} (PURL) or ecosystem-specific package identifiers.
Many records additionally reference alternative identifiers originating
from vendor advisories (e.g., RHSA, DSA, USN), project-specific databases,
or other public vulnerability repositories
\cite{osv2026schema,ghsa2026,redhat2026backport,debian2026security,ubuntu2026security}. Together, these identities form
a network of cross-references that supports the correlation of vulnerability
information across heterogeneous sources
\cite{cpe2026,purl2026,osv2026schema}.
Within a software artifact, software components may appear as binary files,
software packages, container contents, or entries in a software bill of
materials (SBOM). Their reliable mapping therefore requires additional
information such as the package ecosystem, namespace, vendor, version, and
the resolution of alias relationships between different identity models and
advisory records
\cite{qin2023vulnerability,wu2024affected}.

Once the mapping has been established, the available information must be
assessed in relation to the application context, such as the actual use
and reachability of a component and the criticality of the affected
system. The resulting finding therefore combines public vulnerability
information with locally determined component and context data.
Section~\ref{sec:discrepancies} examines the stages of this process at
which differing analysis results may arise.

%%%%%%%%%%%%%%%%%%%%%%%%%%%%%%%%%%%%%%%%%%%%%%%%%%%%%%%%%%%%%%%%%%%%%%%%%%%%%%%%
\section{Causes of Divergent Analysis Results}
\label{sec:discrepancies}
%%%%%%%%%%%%%%%%%%%%%%%%%%%%%%%%%%%%%%%%%%%%%%%%%%%%%%%%%%%%%%%%%%%%%%%%%%%%%%%%

The processing described in the previous section involves several
transformation and decision steps. Differences between analysis results may
already arise from the selection and current state of the information
sources used. Further discrepancies can result from identity and version
resolution or from the final assessment within a particular application
context. The following subsections examine these causes in more detail.

%%%%%%%%%%%%%%%%%%%%%%%%%%%%%%%%%%%%%%%%%%%%%%%%%%%%%%%%%%%%%%%%%%%%%%%%%%%%%%%%
\subsection{Heterogeneous Information Sources}
%%%%%%%%%%%%%%%%%%%%%%%%%%%%%%%%%%%%%%%%%%%%%%%%%%%%%%%%%%%%%%%%%%%%%%%%%%%%%%%%

Although many publicly available information sources use the same
vulnerability identifiers, they differ in purpose, scope, and the metadata
they provide. These differences reflect the distinct roles that the sources
perform within the open-source vulnerability ecosystem. Some adopt
information from primary sources, while others supplement or reassess it.
Differences between analysis results may therefore originate in the
available knowledge base rather than in the analysis of the software system
itself. Previous research has shown that public vulnerability information
can vary in completeness, consistency, and structure
\cite{dong2019viem}.

The heterogeneity concerns both the content and the time at which
information becomes available. Some information systems focus on
standardized descriptions and assessments, whereas others provide detailed
information about package versions, project-specific characteristics, or
distribution-specific modifications. For CVE-2021-44228
(\emph{Log4Shell}), for example, CVE and NVD provided a general description
and assessment, while package databases and project advisories supplied
more precise information about Maven versions, updates, and mitigation
measures. Vendor and project advisories may also publish information about
backports, workarounds, or recommended actions before these details have
been fully incorporated into public databases. Regional systems add
national or regulatory perspectives to the information available from
global sources. Several internally consistent but differently detailed and
time-dependent representations of the same vulnerability may therefore
coexist.

For automated analyses, this means that identical software artifacts may
produce different results solely because different information sources are
used. The differences may concern not only the number of vulnerabilities
identified but also the version ranges classified as vulnerable,
additional references, or severity assessments. The integration of
heterogeneous information therefore remains a central challenge for
current vulnerability management systems and continues to be addressed in
research \cite{qin2023vulnerability,ruan2024vulzoo}. Even where the
underlying information is identical, however, differences may arise from
the identity and version models used for mapping software components.

%%%%%%%%%%%%%%%%%%%%%%%%%%%%%%%%%%%%%%%%%%%%%%%%%%%%%%%%%%%%%%%%%%%%%%%%%%%%%%%%
\subsection{Identity and Version Models}
%%%%%%%%%%%%%%%%%%%%%%%%%%%%%%%%%%%%%%%%%%%%%%%%%%%%%%%%%%%%%%%%%%%%%%%%%%%%%%%%

The mapping of vulnerability information to software components described
in Section~\ref{subsec:interpretation} may produce different results
depending on the identity model used. Public information systems, package
ecosystems, and software artifacts use their own namespaces, identifiers,
and version conventions. No consistently unambiguous mapping exists
between these representations.
The same Log4j component may, for example, appear in a software artifact as
\texttt{log4j-core-2.14.1.jar},
in the Maven ecosystem as
\texttt{org.apache.logging.log4j:log4j-core:2.14.1},
and in a product-oriented database as the CPE name
\texttt{cpe:2.3:a:apache:log4j:2.14.1:*:*:*:*:*:*:*}. A correct analysis
requires these representations to be recognized as referring to the same
software component. If the connection is missing or a name is assigned to
the wrong component, the resulting findings will differ.

Further difficulties arise from project renaming, forks,
distribution-specific packages, and inconsistently maintained namespaces.
Version schemes are also not always directly comparable. A numerically
higher version number does not necessarily indicate a more recent security
state when fixes have been backported to an older release
\cite{redhat2026backport}. A purely numerical comparison may therefore
classify a maintained distribution version as vulnerable even though the
relevant correction has already been applied.

Empirical studies indicate that the reliable mapping of vulnerability
information to the software components actually in use remains one of the
main challenges in automated vulnerability analysis. Incorrect or
incomplete identity resolution can produce both false-positive and
false-negative results and directly affects subsequent assessments.
Current research therefore examines methods for identifying software
components more precisely and linking them across package ecosystems and
information models \cite{wu2024affected,zhao2023sca}. Even where the
mapping is correct, however, analysis results remain dependent on the time
of analysis.

%%%%%%%%%%%%%%%%%%%%%%%%%%%%%%%%%%%%%%%%%%%%%%%%%%%%%%%%%%%%%%%%%%%%%%%%%%%%%%%%
\subsection{Temporal Dynamics}
%%%%%%%%%%%%%%%%%%%%%%%%%%%%%%%%%%%%%%%%%%%%%%%%%%%%%%%%%%%%%%%%%%%%%%%%%%%%%%%%

Vulnerability information is not static but develops throughout its
lifecycle. After initial publication, descriptions may be extended,
version ranges refined, references added, and assessments revised. A
vulnerability analysis therefore represents the state of information at a
particular point in time. Even when the same software components are
examined, analyses performed at different times may produce different
results \cite{nvd2022process,mandl2026groundtruth}.

This dynamic also results from information systems maintaining and updating
their data independently. New findings do not become available in all
sources at the same time. Vendor and project advisories may appear before
their contents have been fully incorporated into public databases, while
downstream systems may add assessments, references, or version information
only gradually. Temporary differences in the available information may
therefore occur, and the same vulnerability may be described differently
depending on the source and the time of analysis
\cite{dong2019viem,nvd2022process}.

Changes over time concern not only the description of a vulnerability but
also its assessment. Further investigation may reveal additional
information about relevant versions, practical exploitability, or available
countermeasures. Such information may alter the priority assigned to a
finding. Vulnerability management is therefore not a one-time analysis but
requires the information base to be updated and reassessed continuously
\cite{nist2022patch,mandl2026groundtruth}. Even when the information is
current and consistent, the final result still depends on the application
context.

%%%%%%%%%%%%%%%%%%%%%%%%%%%%%%%%%%%%%%%%%%%%%%%%%%%%%%%%%%%%%%%%%%%%%%%%%%%%%%%%
\subsection{Context-Dependent Interpretation}
\label{subsec:context}
%%%%%%%%%%%%%%%%%%%%%%%%%%%%%%%%%%%%%%%%%%%%%%%%%%%%%%%%%%%%%%%%%%%%%%%%%%%%%%%%

Even when identical vulnerability information is available at the same
time and the software components have been mapped unambiguously, the
results must still be assessed within the relevant application context.
The risk associated with a vulnerability depends on the properties of the
specific system, its operating environment, and the protective measures in
place. The same vulnerability may therefore receive different priorities
in different organizations or applications despite an identical
information base \cite{nist2022patch}.

The technical severity of a vulnerability can initially be expressed using
the \emph{Common Vulnerability Scoring System} (CVSS)
\cite{cvss40}. CVSS assigns a severity score on a scale from 0.0 to 10.0,
where higher values indicate a greater technical impact. For example,
Log4Shell (CVE-2021-44228) received the maximum base score of 10.0
(\emph{Critical}) because it enables unauthenticated remote code execution
under common deployment conditions. This score, however, does not by
itself determine the significance of the vulnerability for a specific
software system.

Further information may therefore be included in a risk-oriented
prioritization. The \emph{Known Exploited Vulnerabilities Catalog} (KEV)
identifies vulnerabilities for which active exploitation has been observed
in practice \cite{cisa2026kev}. A typical KEV entry contains the CVE
identifier, the affected vendor and product, the date on which the
vulnerability was added to the catalog, and a required remediation action.
For example, the Log4Shell vulnerability (CVE-2021-44228) is listed in the
KEV catalog together with the recommendation to apply the vendor-provided
security update.

The \emph{Exploit Prediction Scoring System} (EPSS) estimates the
probability that a vulnerability will be exploited within the following
30 days by assigning a score between 0 and 1, where higher values indicate
a greater likelihood of exploitation
\cite{first2026epss}. For example, a vulnerability with an EPSS score of
0.95 is considered substantially more likely to be exploited than one
with a score of 0.02, even if both have the same CVSS severity.

The \emph{Vulnerability Exploitability eXchange} (VEX) is a
machine-readable format that allows software suppliers to communicate
whether and under which conditions a known vulnerability affects a
specific product \cite{cyclonedx2026vex}. A typical VEX statement
references a vulnerability (e.g., CVE-2021-44228), the affected product,
an exploitability status such as \emph{affected}, \emph{not affected}, or
\emph{under investigation}, and a corresponding justification. For
example, a supplier may state that a product contains the vulnerable
Log4j library but that the vulnerable code is not reachable in the
delivered configuration. Such a statement allows organizations to
distinguish between a vulnerable component that requires remediation and
one that does not pose an exploitable risk in the given deployment.

A meaningful assessment also requires knowledge about the specific system
that cannot be obtained from general vulnerability databases. Relevant
information includes the actual use and reachability of a software
component within an execution path, existing protective and compensating
controls, the criticality of the supported business process, and internal
security policies \cite{ponta2020detection,nist2022patch}. Analysis tools
differ in the information they consider and in the rules by which they
derive a final assessment.

Figure~\ref{fig:figure_4} illustrates a possible decision process for the
context-dependent treatment of a vulnerability finding. The process begins
by verifying that the software component is present in the artifact. It
then considers reachability and the availability of a fix or workaround.
Depending on the result, the finding may be dismissed based on a VEX
statement, mitigated through a technical measure, or accepted as a
documented risk. Differences between analysis results therefore do not
necessarily indicate incorrect data. They may reflect different technical
assumptions, contextual information, or prioritization strategies.
Empirical comparisons of current software composition analysis tools
support this observation, as the tools may produce substantially
different results even when they examine the same artifacts
\cite{imtiaz2021comparative,mandl2026groundtruth}.

%%%%%%%%%%%%%%%%%%%%%%%%%%%%%%%%%%%%%%%%%%%%%%%%%%%%%%%%%%%%%%%%%%%%%%%%%%%%%%%%
\subsection{Summary}
%%%%%%%%%%%%%%%%%%%%%%%%%%%%%%%%%%%%%%%%%%%%%%%%%%%%%%%%%%%%%%%%%%%%%%%%%%%%%%%%

The causes discussed in this section rarely occur in isolation.
Heterogeneous information sources, different identity and version models,
changes over time, and context-dependent interpretation influence one
another and jointly determine the outcome of a vulnerability analysis.
They give rise to technical challenges for vulnerability management
systems as well as further questions for research and practice, which are
discussed in the following section.

%%%%%%%%%%%%%%%%%%%%%%%%%%%%%%%%%%%%%%%%%%%%%%%%%%%%%%%%%%%%%%%%%%%%%%%%%%%%%%%%
\section{Implications for Research and Practice}
\label{sec:implications}
%%%%%%%%%%%%%%%%%%%%%%%%%%%%%%%%%%%%%%%%%%%%%%%%%%%%%%%%%%%%%%%%%%%%%%%%%%%%%%%%

The preceding sections have shown that divergent analysis results cannot be
attributed to individual tools or databases alone. They arise from the
interaction of several causes. Differences between results should therefore
not be viewed solely as a quality problem of particular tools, but also as a
consequence of a distributed information ecosystem whose contents continue
to change.

This perspective has implications for both academic research and the
practical use of automated vulnerability management systems. Research must
address the integration, consistency, and comparability of heterogeneous
vulnerability information. In practice, the main challenge lies in
establishing reliable processes for handling distributed and changing
vulnerability knowledge.

%%%%%%%%%%%%%%%%%%%%%%%%%%%%%%%%%%%%%%%%%%%%%%%%%%%%%%%%%%%%%%%%%%%%%%%%%%%%%%%%
\subsection{Implications for Research}
%%%%%%%%%%%%%%%%%%%%%%%%%%%%%%%%%%%%%%%%%%%%%%%%%%%%%%%%%%%%%%%%%%%%%%%%%%%%%%%%

The analysis of the open-source vulnerability ecosystem suggests that
future research should examine the complete information flow rather than
isolated data sources or analysis tools. Many comparative studies assess
differences between tools on the basis of their output without
systematically considering the information sources, identity models, and
data versions underlying the respective results. Claims about the quality
of a method are therefore only partly comparable
\cite{imtiaz2021comparative,mandl2026groundtruth}.

Reproducible evaluation methods remain another important research topic.
Because vulnerability information is updated continuously, future studies
should document the state of the information used, the time of analysis,
and the underlying data sources. This documentation is necessary for
meaningful comparisons and long-term reproducibility. Reliable reference
datasets are equally important. Such datasets should combine different
identity models, versions, and contextual information rather than relying
exclusively on individual databases or CVE lists
\cite{mandl2026groundtruth,nvd2022process}.

AI-assisted methods may contribute to this work. Named-entity recognition
and relation extraction can identify vendor, product, and version
information in unstructured advisories and transfer it into common data
models. Recent approaches combine these methods with unified identity
schemas and graph-based models to resolve inconsistent names and
configuration-dependent relationships
\cite{jiang2025vulcpe}. Embedding methods and large language models may
also help identify semantically related product names, connect aliases, and
interpret heterogeneous descriptions
\cite{sheng2025llmsecurity}. Hybrid methods appear particularly suitable
for this purpose, as they combine machine-generated mappings with formal
identifiers, version rules, and traceable source references. Their
evaluation should consider not only mapping accuracy but also
interpretability, stability over time, and the provenance of the
information used.

The ecosystem model presented in this paper also indicates that identity
resolution and information integration remain open research problems.
Consistent mappings between product-oriented and package-oriented identity
models are still difficult to establish. The same applies to the inclusion
of additional contextual information. Improvements in these areas could
make differences between analysis results easier to explain and could
increase the reliability of the resulting assessments.

%%%%%%%%%%%%%%%%%%%%%%%%%%%%%%%%%%%%%%%%%%%%%%%%%%%%%%%%%%%%%%%%%%%%%%%%%%%%%%%%
\subsection{Implications for Practice}
%%%%%%%%%%%%%%%%%%%%%%%%%%%%%%%%%%%%%%%%%%%%%%%%%%%%%%%%%%%%%%%%%%%%%%%%%%%%%%%%

A Software Bill of Materials (SBOM) is a formal record of the components
contained in software and their relationships within the software supply
chain \cite{ntia2021sbom}. In practice, the ecosystem model presented here
means that vulnerability management should not be reduced to matching an
SBOM against a single vulnerability database. Vulnerability information
originates from different sources, changes over time, and must be
interpreted within a particular application context. Analysis results
therefore represent a technically justified view of the information
available at a given time. Differences between tools or analysis dates are
to be expected and should be taken into account when results are assessed.

Organizations should not adopt vulnerability information without examining
its origin, currency, and technical interpretation. Documenting the
information sources used, the time of analysis, and the software artifacts
examined makes decisions easier to trace and repeated analyses easier to
compare. Regular reassessment is also necessary because version mappings,
technical assessments, and contextual information may change during the
lifecycle of a vulnerability
\cite{nvd2022process,nist2022patch,mandl2026groundtruth}.

\begin{figure*}[t]
    \centering
    \includegraphics[
        width=\textwidth,
        trim=0 70 0 70,
        clip
    ]{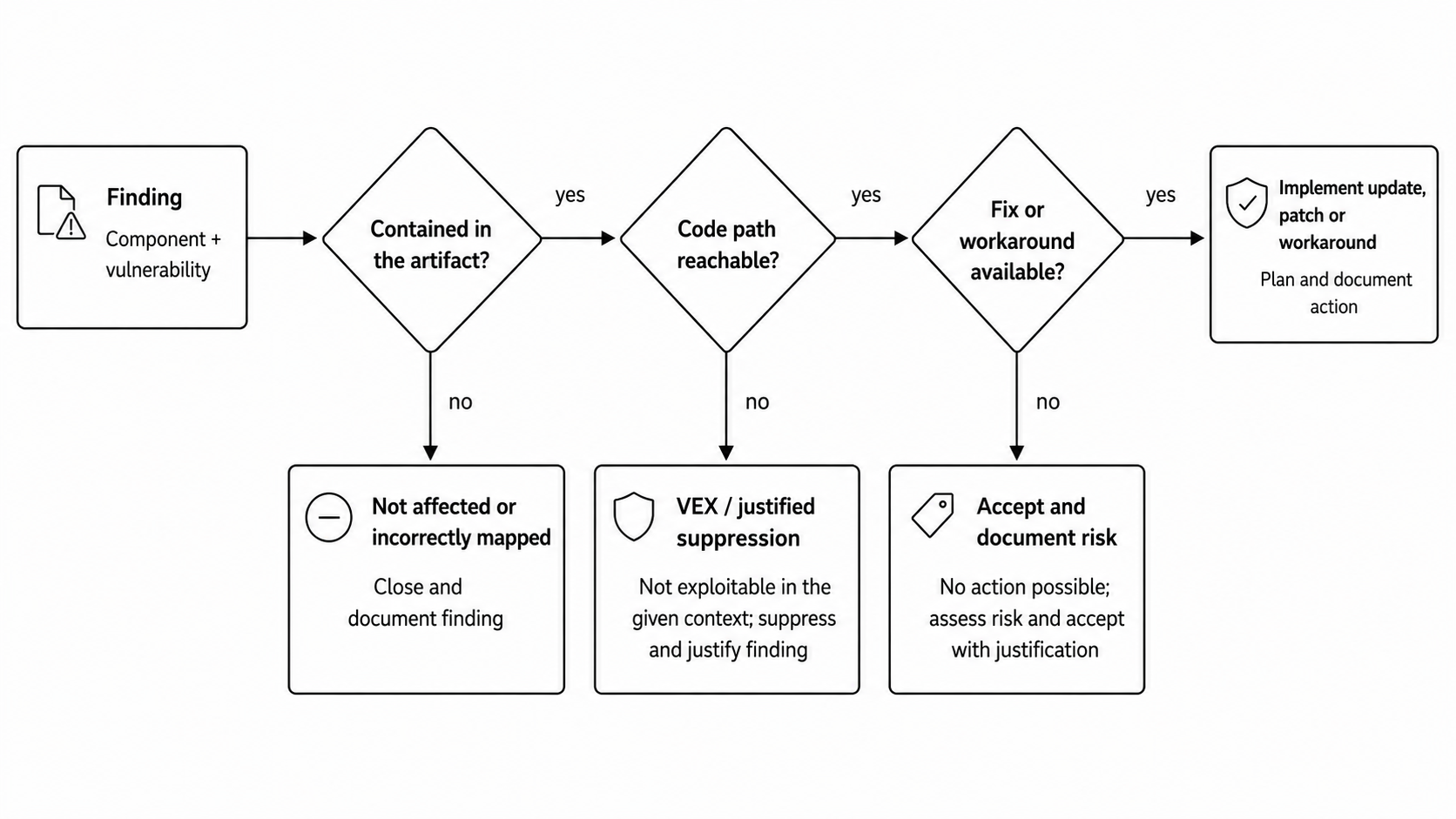}
    \caption{Context-dependent assessment and treatment of a vulnerability
    finding. The decision process extends from validating the finding and
    prioritizing it in light of contextual information to its technical or
    organizational treatment.}
    \label{fig:figure_4}
\end{figure*}

Technical vulnerability information alone does not provide a sufficient
basis for sound prioritization. A risk-oriented assessment additionally
requires the local context set out in Section~\ref{subsec:context}, such
as how a component is actually used, whether it is reachable, and how
critical the supported business process is. Information from CVSS, KEV,
EPSS, or VEX may support this process, but it cannot replace an
application-specific assessment
\cite{cvss40,cisa2026kev,first2026epss,cyclonedx2026vex,
nist2022patch}. Figure~\ref{fig:figure_4} shows the subsequent decision
process, from validating a finding to its technical or organizational
treatment.

The decisive factor is therefore not the selection of a single data source
or analysis tool, but the ability to combine heterogeneous vulnerability
information consistently, keep it current, and interpret it within the
relevant application context. The ecosystem model developed in this paper
provides a conceptual framework for this task. It also supports the
interpretation of divergent analysis results as consequences of a
distributed information and decision process.

%%%%%%%%%%%%%%%%%%%%%%%%%%%%%%%%%%%%%%%%%%%%%%%%%%%%%%%%%%%%%%%%%%%%%%%%%%%%%%%%
\section{Conclusion}
\label{sec:conclusion}
%%%%%%%%%%%%%%%%%%%%%%%%%%%%%%%%%%%%%%%%%%%%%%%%%%%%%%%%%%%%%%%%%%%%%%%%%%%%%%%%

Free and open-source software forms an essential part of modern software
systems. As the number of public vulnerability information sources, package
ecosystems, and analysis tools has grown, their evaluation has become more
complex. Divergent analysis results should therefore not be regarded as
exceptional, but as a structural property of a distributed vulnerability
ecosystem that changes continuously.

This paper has described the open-source vulnerability ecosystem as a
continuous information flow. The focus was not on individual databases or
tools, but on the origin, integration, transformation, and
context-dependent interpretation of vulnerability information. On this
basis, heterogeneous information sources, differing identity and version
models, changes over time, and context-dependent assessment were identified
as major causes of divergent analysis results.

The findings show that differences between analysis tools do not
necessarily indicate incorrect or incomplete findings. They may also result
from differences in the available information, mapping rules, and technical
assumptions. Assessments of automated analyses should therefore take into
account the information sources used, the time of analysis, and the
relevant application context.

Future research should improve the interoperability, traceability, and
reproducibility of distributed vulnerability information. Reliable methods
for identity resolution, consistent exchange models, and closer
consideration of local context are particularly important. AI-assisted methods may contribute to this process by extracting information from unstructured advisories and linking semantically related product and package names. Their outputs, however, should be validated against formal identifiers and version rules to ensure reliable vulnerability matching \cite{jiang2025vulcpe,sheng2025llmsecurity}. In practice, vulnerability
management should not be reduced to matching software against a single data
source, but treated as a continuous information and decision process.

%%%%%%%%%%%%%%%%%%%%%%%%%%%%%%%%%%%%%%%%%%%%%%%%%%%%%%%%%%%%%%%%%%%%%%%%%%%%%%%%
\section*{Note on the Preparation of This Paper}
%%%%%%%%%%%%%%%%%%%%%%%%%%%%%%%%%%%%%%%%%%%%%%%%%%%%%%%%%%%%%%%%%%%%%%%%%%%%%%%%

Large language model tools were used to support individual editorial tasks
and the preparation of the figures. The authors retain full responsibility
for the content, the selection and verification of sources, the
argumentation, and the final version of the paper.

%%%%%%%%%%%%%%%%%%%%%%%%%%%%%%%%%%%%%%%%%%%%%%%%%%%%%%%%%%%%%%%%%%%%%%%%%%%%%%%%
\appendix
%%%%%%%%%%%%%%%%%%%%%%%%%%%%%%%%%%%%%%%%%%%%%%%%%%%%%%%%%%%%%%%%%%%%%%%%%%%%%%%%

%%%%%%%%%%%%%%%%%%%%%%%%%%%%%%%%%%%%%%%%%%%%%%%%%%%%%%%%%%%%%%%%%%%%%%%%%%%%%%%%
\section{List of Abbreviations}
\label{app:abbreviations}
%%%%%%%%%%%%%%%%%%%%%%%%%%%%%%%%%%%%%%%%%%%%%%%%%%%%%%%%%%%%%%%%%%%%%%%%%%%%%%%%

Table~\ref{tab:abbreviations} summarizes the abbreviations used frequently
throughout this paper.

\begingroup
\footnotesize
\setlength{\tabcolsep}{4pt}
\renewcommand{\arraystretch}{1.15}

\begin{xltabular}{\linewidth}{L{20mm}L{52mm}Y}
\caption{Abbreviations used in the open-source vulnerability ecosystem}
\label{tab:abbreviations}\\

\toprule
\rowcolor{gray!15}
\textbf{Abbreviation} &
\textbf{Full term} &
\textbf{Meaning} \\
\midrule
\endfirsthead

\caption[]{Abbreviations used in the open-source vulnerability ecosystem
(continued)}\\

\toprule
\rowcolor{gray!15}
\textbf{Abbreviation} &
\textbf{Full term} &
\textbf{Meaning} \\
\midrule
\endhead

\midrule
\multicolumn{3}{r}{\footnotesize Continued on the next page}\\
\endfoot

\bottomrule
\endlastfoot

CNA &
CVE Numbering Authority &
An authorized organization that assigns CVE identifiers. \\

\rowcolor{gray!4}
CNNVD &
China National Information Security Vulnerability Database &
A national vulnerability database operated in China. \\

CNVD &
China National Vulnerability Database &
A national vulnerability database operated in China. \\

\rowcolor{gray!4}
CPE &
Common Platform Enumeration &
A standard for identifying software products. \\

CVE &
Common Vulnerabilities and Exposures &
A globally unique identification system for published vulnerabilities. \\

\rowcolor{gray!4}
CVSS &
Common Vulnerability Scoring System &
A standardized system for assessing the technical severity of a
vulnerability. \\

CWE &
Common Weakness Enumeration &
A classification of software weakness categories. \\

\rowcolor{gray!4}
EPSS &
Exploit Prediction Scoring System &
A model for estimating the probability that a vulnerability will be
exploited in the future. \\

EUVD &
European Vulnerability Database &
The European vulnerability database. \\

\rowcolor{gray!4}
GHSA &
GitHub Security Advisory &
Advisory identifier used in the GitHub Advisory Database. \\

JVN &
Japan Vulnerability Notes &
A Japanese coordinated vulnerability portal; JVN iPedia is the associated
database. \\

\rowcolor{gray!4}
KEV &
Known Exploited Vulnerabilities &
A catalog of vulnerabilities known to be actively exploited. \\

NVD &
National Vulnerability Database &
An information system that enriches CVE records with structured metadata. \\

\rowcolor{gray!4}
OSV &
Open Source Vulnerability Format &
A package-oriented schema and database for vulnerabilities in open-source
software. \\

PSIRT &
Product Security Incident Response Team &
A vendor organization responsible for coordinating, assessing, and
publishing information about vulnerabilities in its products. \\

\rowcolor{gray!4}
PURL &
Package URL &
A standardized identifier for software packages. \\

SBOM &
Software Bill of Materials &
A machine-readable description of the components contained in a software
artifact. \\

\rowcolor{gray!4}
SCA &
Software Composition Analysis &
The analysis of software dependencies for known vulnerabilities. \\

VEX &
Vulnerability Exploitability eXchange &
A machine-readable statement about whether a vulnerability can be
exploited within a particular product. 
\end{xltabular}
\endgroup

%%%%%%%%%%%%%%%%%%%%%%%%%%%%%%%%%%%%%%%%%%%%%%%%%%%%%%%%%%%%%%%%%%%%%%%%%%%%%%%%

\end{document}